\documentstyle[prl,aps,epsf,twocolumn]{revtex}
\begin{document}
\draft
\twocolumn[\hsize\textwidth\columnwidth\hsize\csname@twocolumnfalse%
\endcsname

\preprint{}

\title{Confinement of Slave-Particles in $U(1)$
Gauge Theories of Strongly-Interacting Electrons}
\author{Chetan Nayak}
\address{Physics Department, University of California, Los Angeles, CA 
  90095--1547}
\date{\today}
\maketitle

\begin{abstract}
We show that slave particles are always confined
in $U(1)$ gauge theories of interacting electron systems.
Consequently, the low-lying degrees
of freedom are different from the slave particles.
This is done by constructing a
dual formulation of the slave-particle
representation in which the
no-double occupany constraint becomes linear
and, hence, soluble. Spin-charge separation,
if it occurs, is due to the existence
of solitons with fractional quantum numbers.
\end{abstract}
\vspace{1 cm}

]
\narrowtext

{\it Introduction.} A number of
attempts \cite{Baskaran87,Baskaran88,Kotliar88,Ioffe89,Lee92,Tikofsky94,Altshuler96,Wen96,Kim99} to understand the
dynamics of strongly-correlated electrons
have employed a `slave particle' decomposition of
the electron operator, such as
\begin{equation}
\label{eqn:slave_decomp}
{c_{i\sigma}} = {b_i^\dagger}{f_{i\sigma}}
\end{equation}
The slave boson operator, ${b_i}$, is supposed to carry the
charge of the electron while the fermion
${f_{i\sigma}}$ carries the spin.
It is sometimes supposed that these particles
can be liberated at low energies, with
spin-charge separation as an  upshot.

In order for this to occur, the $U(1)$
gauge symmetry
\begin{eqnarray}
\label{eqn:gauge_symm1}
{b_i} \rightarrow {e^{i\theta_i}}{b_i},
{\hskip 0.7 cm}
{f_{i\sigma}} \rightarrow {e^{i\theta_i}}{f_{i\sigma}}
\end{eqnarray}
must be broken.This $U(1)$, it should be emphasized,
is {\it not} the electromagnetic $U(1)$. It is due to
the redundancy inherent in the slave particle
description (\ref{eqn:slave_decomp}).
It will be broken if
the gauge field which parametrizes fluctuations about the
broken symmetry state is not confining.
Such a gauge theory is said to be in its Coulomb
phase. In this paper, we suggest that this
cannot happen. Since the gauge symmetry
(\ref{eqn:gauge_symm1}) is an exact {\it local} symmetry of the model
it can never be broken.
This follows from Elitzur's theorem \cite{Elitzur75}.
Global symmetries can be broken since the effect of an infinitesimal
symmtery-breaking field on long-wavelength fluctuations
is extensive; in the infinite-volume
limit, it can inhibit the symmetry-restoring
fluctuations. Local symmetries cannot be broken
since they can be restored by purely local (pure gauge)
fluctuations. The effects of an infinitesimal symmetry-breaking
field are infinitesimal, in contrast to the
global case; in this respect, gauge theories are similar
to $1D$ systems.

Consequently, perturbation theory about an assumed
broken-symmetry state is not valid. The
slave-particle gauge theory is infinitely
strongly-coupled -- i.e. there is no kinetic energy
for the gauge field. In this paper, we will discuss a way of analyzing
the strongly-coupled theory and ramifications
for the issue of spin-charge separation in strongly-correlated
electron systems.

{\it $U(1)$ Gauge Theory Formulation of the $t-J$ Model.}
It is often convenient to reformulate models
of strongly-interacting electrons in terms of
redundant auxiliary degrees of freedom which satisfy
constraints reducing their enlarged Hilbert
space to the physical one. Consider, for instance,
the Hubbard model in the limit that the on-site repulsion
is much larger than the hopping matric element, $U>>t$.
At energy and temperature
scales much less than $U$, we may replace this
model by the $t-J$ model
\begin{equation}
H = -t {\sum_{<i,j>}}\left({c^\dagger_{i\sigma}}{c^{}_{j\sigma}}
+ {\rm h.c.}\right)
+ J {\sum_{<i,j>}} \left({{\bf S}_i}\cdot{{\bf S}_j}
-\frac{1}{4}{n_i}{n_j}\right)
\end{equation}
with $J={t^2}/4U$ {\it together with
the constraint} ${n_i}\leq 1$.
The constraint is an extreme form
of strong interactions; it makes the physics
of the $t-J$ model opaque. By contrast,
the residual interaction, $J$,
is small.

We can rewrite this Hamiltonian using the slave
particle representation (\ref{eqn:slave_decomp}).
A slave boson ${b_i^\dagger}|0\rangle$ represents
a vacant site, while an auxiliary fermion
${f^\dagger_{i\sigma}}|0\rangle$ represent a site
occupied by an electron of spin $\sigma$. The
no-double-occupancy constraint ${n_i}\leq 1$
now reads
\begin{equation}
{b_i^\dagger}{b^{}_i} + {f^\dagger_{i\sigma}}{f^{}_{i\sigma}}
= 1
\end{equation}
It restricts the Hilbert space of slave bosons
and auxiliary fermions to the physical subspace which
only has the above three states per site. Note
that the constraint is now an equality rather than
an inequality, thereby allowing a Lagrange
multiplier formulation.

The Hamiltonian can now be written
\begin{eqnarray}
H &=& -t {\sum_{<i,j>}}{f^\dagger_{i\sigma}}{f^{}_{j\sigma}}
{b^{}_i}{b^\dagger_j} + {\rm h.c.}
- 2J {\sum_{<i,j>}} {f^\dagger_{i\sigma}}{f^{}_{j\sigma}}
{f^\dagger_{j\alpha}}{f_{i\alpha}}\cr & &\,\,
+ {\sum_i} {a^0_i}\left({b_i^\dagger}{b_i} +
{f^\dagger_{i\sigma}}{f^{}_{i\sigma}} - 1\right)
\end{eqnarray}
where $a^0$ is the Lagrange multiplier which
enforces the constraint. Following \cite{Baskaran87,Marston89}, we
now decouple the quartic terms
with the aid of a Hubbard-Stratonovich
field ${\chi_{ij}}$.
\begin{eqnarray}
H &=& {\sum_{<i,j>}} \biggl(J{\left|{\chi_{ij}}\right|^2}
- J {\chi_{ij}}{f^\dagger_{i\sigma}}{f^{}_{j\sigma}} + {\rm h.c.}
\cr & &{\hskip 1 cm}
- t\, {\chi_{ij}}{b^\dagger_j}{b^{}_i} + {\rm h.c.}
- \frac{t^2}{J} 
{b^\dagger_i}{b^{}_i}{b^\dagger_j}{b^{}_j}\biggr)\cr & &
+ {\sum_i} {a^0_i}\left({b_i^\dagger}{b_i} +
{f^\dagger_{i\sigma}}{f^{}_{i\sigma}} - 1\right)
\end{eqnarray}
When ${\chi_{ij}}$ aquires an expectation value,
the symmetry (\ref{eqn:gauge_symm1}) is broken.
We will assume that the magnitude of ${\chi_{ij}}$
is fixed and study fluctuations of the phase
\begin{equation}
{\chi_{j,j+{\bf k}}} = \chi \, {e^{i {a_k}(j)}}
\end{equation}
When these phase fluctuations are large,
the symmetry is restored.
In the continuum limit, we can write the corresponding
Lagrangian as \cite{Ioffe89,Lee92}:
\begin{eqnarray}
{\cal L} &=&  {f^\dagger_{\sigma}}
\left({\partial_\tau} - {a_0}\right){f^{}_{\sigma}} +
{f^\dagger_{\sigma}}
\frac{1}{m_s}{\left(i\vec{\nabla} - \vec{a}\right)^2}
{f^{}_{\sigma}} -\, {a_0}{\rho_0}\cr & & +\,
{b^\dagger}\left({\partial_\tau} - {a_0} - {A_0}\right )b +
{b^\dagger}
\frac{1}{m_c}{\left(i\vec{\nabla} - \vec{a} - \vec{A}\right)^2}
{b} + {|b|^4}
\end{eqnarray}
where ${m_s}\sim 1/J$, ${m_c}\sim 1/t$, and ${\rho_0}=1/{a^d}$,
where $a$ is the lattice spacing.
In this Lagrangian, we have explicitly written the coupling
to the external electromagnetic field. We have coupled it
to the slave bosons, but this is purely a matter of taste.
We could have coupled $\vec{A}$ to the fermions
instead; the physics would be the same thanks to the
constraint. This arbitrariness is a reflection of the fact
that the slave particles are confined, as we will
see later. We may take $A_0$ to include the chemical
potential; by varying it, we can change the electron density
and, hence, the fermion and boson densities.
The gauge symmetry of this model,
\begin{eqnarray}
\label{eqn:gauge}
{b}(x) &\rightarrow& {e^{i\theta(x)}}{b}(x)\cr
{f_\sigma}(x) &\rightarrow& {e^{i\theta(x)}}{f_\sigma}(x)\cr
{a_\mu}(x) &\rightarrow& {a_\mu}(x) - {\partial_\mu}\theta(x)
\end{eqnarray}
reflects the redundancy of the slave-particle
description.

Notice that there is no kinetic term ${f^2_{\mu\nu}}$
for the gauge field $a_\mu$. Hence, this is a theory
of spin-$1/2$ fermions and spinless bosons
interacting with a gauge field {\it at infinite
coupling}. The time component of the
gauge field, $a_0$, simply enforces the
no-double occupancy constraint, thereby reducing the
redundancy of the slave-boson representation.
The spatial components, $a_i$, simply restore
the broken gauge symmetry (\ref{eqn:gauge_symm1}).
Since the gauge field is at infinite coupling,
it is necessarily a confining gauge field: all physical
states are gauge singlets and the symmetry
(\ref{eqn:gauge_symm1}) is restored. In particular, the slave
bosons, $b$, and the fermions, ${f^{}_{\sigma}}$
are not part of the physical spectrum. Nevertheless,
they
have been treated as quasiparticles
{\it weakly-coupled} to a gauge field
\cite{Baskaran87,Baskaran88,Kotliar88,Ioffe89,Lee92,Tikofsky94,Altshuler96,Wen96,Kim99}. A safer way of proceeding is by solving the
constraints which follow from
integrating out the gauge field $a_\mu$.
In this paper, we show how this can be done.

{\it Slave Particle Confinement in $1D$.}
The basic strategy can be demonstrated in
the $1+1$-dimensional case. For simplicity,
we consider the case of $1/2$-filling, at which
there are no slave bosons. Using bosonization,
we rewrite the fermions, ${f^{}_{\sigma}}$, in terms
of bosonic fields $\phi_f$ and $\phi_s$:
\begin{eqnarray}
\label{eqn:bose_slaves}
{f^{}_{R,L;\sigma}} = {e^{\frac{i}{\sqrt{2}}{\phi^{R,L}_f}}}
{e^{\frac{i}{\sqrt{2}}\sigma{\phi^{R,L}_s}}}
\end{eqnarray}
where ${f^{}_{\sigma}} = {e^{-i{k_F}x}}\,\,{f^{}_{R;\sigma}}
+ {e^{i{k_F}x}}\,\,{f^{}_{L;\sigma}}$.
There are two bosonic fields, ${\phi_f}$ and ${\phi_s}$,
because the fermions ${f^{}_{\sigma}}$ carry two quantum 
numbers, fermion number and charge. The gauge field,
$a_\mu$, couples to the fermion number. The fermons
are neutral, so they are not coupled to the
electromagnetic gauge field.
The Lagrangian can now be rewritten
in terms of ${\phi_s}={\phi^R_s}+{\phi^L_s}$
and the dual scalar field, ${{\tilde \phi}_f}$,
defined by
\begin{eqnarray}
{\partial_\mu}{{\tilde \phi}_f} =
{\epsilon_{\mu\nu}}\left({\partial_\nu}{\phi_f} -
{a_\nu}\right)
\end{eqnarray}
In these variables, it is:
\begin{eqnarray}
{\cal L} = \frac{1}{8\pi}\,{\left({\partial_\mu}{\phi_s}\right)^2} +
\frac{1}{8\pi}\,{\left({\partial_\mu}{{\tilde \phi}_f}\right)^2}
+ {a_\mu}{\partial_\mu}{{\tilde \phi}_f}
\end{eqnarray}
Notice that the constraint is linear in
the dual bosonized representation. Hence, we can
simply solve it,
\begin{equation}
{\partial_\mu}{{\tilde \phi}_f} = 0
\end{equation}
thereby passing to a Lagrangian which
only contains physical, gauge-neutral variables:
\begin{equation}
{\cal L} = \frac{1}{8\pi}\,{\left({\partial_\mu}{\phi_s}\right)^2}
\end{equation}
The solitons of this Lagrangian are
spinons, neutral spin-$1/2$ excitations created
by ${e^{\frac{i}{\sqrt{2}}\sigma{\phi^{R,L}_s}}}$;
they are clearly not the same as the $f_\sigma$'s,
despite having the same spin
and charge quantum numbers,
as may be seen by comparison with (\ref{eqn:bose_slaves}).
The former are physical, gauge-invariant excitations,
while the latter are not part of the physical spectrum.
A similar result was found by Mudry and Fradkin  \cite{Mudry94}
for the $1+1-D$ $t-J$ model.

The key step in this analysis was the use of
a bosonized representation. Since the constraint is
{\it linear} in this representation, it can be solved,
thereby leaving only the physical degrees of freedom in
the Lagrangian. The purpose of this paper is to do
this in $2+1$-dimensions, a task to
which we turn in the next section.

{\it Slave Particles in $2D$.} In two-dimensions,
we can use boson-vortex duality
\cite{Peskin78,Thomas78,Wen90,Dasgupta81,Fisher89}
to represent the slave bosons, $b$.
In order to represent the auxiliary fermions $f_\sigma$,
we use the construction of \cite{Balents99b}.
We represent the auxiliary
$f_\sigma$ as auxiliary bosons interacting with a
$U(1)$ Chern-Simons gauge field which
attaches flux to $S_z$. A dual representation is then
used for these auxiliary bosons. We omit the
details of this construction for $f_\sigma$;
they are given in \cite{Balents99b}, where this
construction is used for electrons themselves.
The resulting dual Lagrangian is
\begin{eqnarray}
{{\cal L}_{dual}} &=& {\sum_\alpha}
{{\cal L}_{GL}}({\Phi^\alpha},
\frac{1}{2}({a^\rho_\mu}\pm{a^\sigma_\mu})) +
{{\cal L}_{GL}}({\Phi^c},{a^c_\mu})\cr
& &+ \,{{\cal L}_{cs}}({a^\sigma_\mu})
+ {1 \over {2\pi}} A_\mu {\epsilon_{\mu \nu \lambda}} 
\partial_\nu {a^c_\lambda}\cr
& & \,{a_\mu}\left[{\epsilon_{\mu\nu\lambda}}{\partial_\nu}
\left({a^{\rho}_\lambda} + {a^{c}_\lambda}\right)
- {\rho_0}{\delta_{\mu 0}}\right]
\end{eqnarray}
where
\begin{eqnarray}
{J^{\rho,\sigma}_\mu} =
{\epsilon_{\mu\nu\lambda}}{\partial_\nu}
{a^{\rho,\sigma}_\lambda}, {\hskip 0.7 cm}
{J^c_\mu} =
{\epsilon_{\mu\nu\lambda}}{\partial_\nu}
{a^c_\lambda}
\end{eqnarray}
are, respectively, the auxiliary fermion
number current, the $S_z$ current,
and the slave boson number current.
The last of these is equal to the charge current.
The up- and down-spin fermion currents
are given by
${J^{\uparrow,\downarrow}_\mu}~=~({J^{\rho}_\mu}~\pm~{J^{\sigma}_\mu})~/~2$.
${\Phi^\uparrow}$,${\Phi^\downarrow}$, and ${\Phi^c}$
annihilate vortices in these currents.
${{\cal L}_{GL}}$ is given by:
\begin{equation}
{{\cal L}_{GL}}({\Phi^c},{a^c_\mu}) = 
 {1 \over 2} {|(i{\partial_\mu} - {a^c_\mu}){\Phi^c} |^2}
+ V({\Phi^c}) + {1 \over 2} {({f^c_{\mu \nu}})^2}   .
\label{eqn:GL_Lag}
\end{equation}
and a similar expression for
${{\cal L}_{GL}}({\Phi^\alpha},({a^\rho_\mu}\pm{a^\sigma_\mu})/2)$.
The  ``potential'' can be expanded as
$V(\Phi) = r|\Phi|^2 + u|\Phi|^4+\ldots$.

The constraint now reads:
\begin{eqnarray}
{\epsilon_{\mu\nu\lambda}}{\partial_\nu}
\left({a^{\rho}_\lambda} + {a^{c}_\lambda}\right)
- {\rho_0}{\delta_{\mu 0}} = 0
\end{eqnarray}
The solution of the constraint is
\begin{equation}
{a^{c}_\lambda} \equiv
- {a^{\rho}_\lambda} + {\rho_0}{\epsilon_{\lambda j}}{x_j}
\end{equation}
where the $\equiv$ sign means equal up to a gauge
transformation. The physics of this equation is simple:
the auxiliary fermion number and slave boson number
currents are not independent; they are equal and
opposite.

Note that this constraint does not commute with
the Hamiltonian (in the terminology introduced by Dirac,
it is a second-class constraint). Hence, we must also
impose the condition which follows from the commutator
of the Hamiltonian with the constraint. Equivalently,
the constraint removes all of $a^c_\mu$ except for a pure gauge degree
of freedom. Before choosing a gauge and eliminating this
degree of freedom, we must impose the equation which follows from
its variation.
For illustrative purposes, we will solve
this equation under the
assumption that the $Z_2$ symmetry
\begin{equation}
\label{eqn:z_2}
{\Phi_{\uparrow,\downarrow}}\rightarrow
- {\Phi_{\uparrow,\downarrow}}  ,
\end{equation}
is unbroken. In the $Z_2$-broken case, a similar
but slightly different solution is available.
We introduce the fields
$\Phi_\rho$ and $\Phi_\sigma$,
following \cite{Balents99b}:
\begin{equation}
{\Phi_\rho}={\Phi_\uparrow}{\Phi_\downarrow}   ;  \hskip0.7cm
{\Phi_\sigma}={\Phi_\uparrow}{\Phi^\dagger_\downarrow}  ,
\end{equation}
which are the appropriate degrees of freedom
when the $Z_2$ symmetry
is unbroken. The effective Lagrangian for
$Z_2$-symmetric phases is:
\begin{equation}
{{\cal L}_{eff}} = {{\cal L}_\rho} + {{\cal L}_\sigma} 
+ {{\cal L}_{GL}}({\Phi^c},{a^c_\mu}) + {{\cal L}_{int}} + {{\cal L}_{con}} ,
\label{eq:eff_Lag}
\end{equation}
with an auxiliary fermion number sector,
${{\cal L}_\rho}={{\cal L}_{GL}}({\Phi^\rho},{a^\rho_\mu})$
and a spin sector,
\begin{eqnarray}
{\cal L}_\sigma  = {{\cal L}_{GL}}({\Phi^\sigma},{a^\sigma_\mu})
+ i {1 \over {4\pi}} \epsilon_{\mu \nu \lambda}
{a^\sigma_\mu} {\partial_\nu} {a^\sigma_\lambda}
\label{L-spin}
\end{eqnarray}
${{\cal L}_{int}}$ contains (subleading) interactions
between the charge and spin sectors. ${{\cal L}_{con}}$
is the constraint.
The equation which follows from the commutator
of the constraint with the Hamiltonian now takes the form:
\begin{equation}
{\rm Im}\left({\Phi^{c\dagger}}(\partial_\mu -
i{a^\rho_\mu}){\Phi^c}\right)
+ {\rm Im}\left({\Phi^{\rho\dagger}}(\partial_\mu -
i{a^\rho_\mu}){\Phi^\rho}\right) = 0
\end{equation}
This may be solved by taking
${\Phi_\rho}={\Phi^{c\dagger}}$.
The effective Lagrangian which
results from solving the constraints is:
\begin{eqnarray}
\label{eqn:master}
{{\cal L}_{\rm eff}}  &=& 
 {1 \over 2} \,{|(\partial_\mu - i{a^\rho_\mu}){\Phi_\rho}|^2}
+ {r'_\rho}{|{\Phi_\rho}|^2} + {u'_\rho} |\Phi_\rho|^4 \cr
& &\,\,\,\,+ {1 \over 2} {(f^\rho_{\mu \nu})^2}
+ {1 \over {2\pi}} A_\mu {\epsilon_{\mu \nu \lambda}} 
\partial_\nu {a^\rho_\lambda}   \cr
& & {1 \over 2} {|(\partial_\mu -
i{a^\sigma_\mu}){\Phi_\sigma
}|^2}
+ {r_\sigma}{|{\Phi_\sigma}|^2} + u_\sigma |\Phi_\sigma|^4 \cr
& &+ 
{1 \over 2} {(f^\sigma_{\mu \nu})^2}
+ i {1 \over {4\pi}} \epsilon_{\mu \nu \lambda}
{a^\sigma_\mu} {\partial_\nu} {a^\sigma_\lambda}
 + {{\cal L}_{\rm int}}
\end{eqnarray}

This is the main result of this paper: after solving the constraint,
we are left with an effective action of the generic form introduced
in \cite{Balents99b}. A similar conclusion has
recently been reached by D.-H. Lee \cite{LeeDH99}.
This Lagrangian contains no fields which transform under
the gauge symmetry (\ref{eqn:gauge}), so the slave particles
are not part of the theory. As discussed in \cite{Balents99b},
the possibility of spin-charge separation is dependent
on the existence of fractional quantum number solitons in this Lagrangian
(see also \cite{Senthil99,Fradkin91}).

{\it Discussion.} We have seen how
spin-charge separation
does not result from the deconfinement of
slave particles. Spin-charge separation, if it occurs,
is due to the existence of solitons with fractional quantum numbers.
Slave particles are always confined by
a $U(1)$ gauge field which is at infinite coupling
since its purpose is to impose a constraint reducing
the Hilbert space to the physical one.
Attempts to treat the gauge field perturbatively
\cite{Baskaran87,Baskaran88,Kotliar88,Ioffe89,Lee92,Tikofsky94,Altshuler96,Wen96,Kim99}
fail to impose the constraint and therefore lead, incorrectly,
to the conclusion that the slave particles can be deconfined.
The fallacy can be seen by considering:
\begin{equation}
S = \int {d^d}x\,\left(\overline{\psi^\alpha} {\gamma^\mu}
\left(i{\partial_\mu} - {a_\mu}\right) {\psi_\alpha}
+ \overline{\chi} {\gamma^\mu}
\left(i{\partial_\mu} - {a_\mu} - {A_\mu}\right) \chi
\right)
\end{equation}
${\psi_\alpha}$ carries spin but no charge,
while $\chi$ carries charge but no spin.
${A_\mu}$ is the electromagnetic field
while ${a_\mu}$ imposes the constraint
${j_\mu} = \overline{\psi^\alpha} {\gamma_\mu} {\psi_\alpha}
+  \overline{\chi}{\gamma_\mu}\chi = 0$.
This constraint holds at every point in space, at
any scale at which we choose to probe the
system. On the other hand, one might imagine that one
can use an RG transformation to produce a low-energy
effective field theory of the form
\begin{eqnarray}
S &=& \int {d^d}x\,\biggl(\overline{\psi^\alpha} {\gamma^\mu}
\left(i{\partial_\mu} - {a_\mu}\right) {\psi_\alpha}\cr
& &{\hskip 0.5 cm}+ \overline{\chi} {\gamma^\mu}
\left(i{\partial_\mu} - {a_\mu} - {A_\mu}\right)
+ \frac{1}{2e^2}{f_{\mu\nu}^2}\biggr)
\end{eqnarray}
since integrating out the fermions would appear to
generate such a term. One can imagine that
such a theory will have a Coulomb phase,
in which the fermions are weakly-coupled. (Note, however,
that it is believed \cite{Fradkin79} that, even for this
model with finite coupling, the
Coulomb phase can occur only for $D\geq 3$.)
However, this line of reasoning
is incorrect. An RG transformation should integrate out
the short-distance fluctuations of
both the fermions {\it and} the gauge fields. Since the
fermions and gauge fields are infinitely strongly-coupled
at short-distances (i.e. in the bare action),
this procedure cannot be done perturbatively.
The strong-coupling expansion should be used instead, and
it leads to the conclusion that the fermions are
confined. It {\it is} permissible, at least formally,
to integrate out the $\chi$ field alone, to derive an
effective action for the gauge fields and $\psi_\alpha$:
\begin{eqnarray}
S &=& \int {d^d}x\,\left(\overline{\psi^\alpha} {\gamma^\mu}
\left(i{\partial_\mu} - {a_\mu}\right) {\psi_\alpha}
+ \frac{1}{2e^2}{{\cal L}_M}(a+A)\right)\cr
&=& \int {d^d}x\,\left(\overline{\psi^\alpha} {\gamma^\mu}
\left(i{\partial_\mu} - {a'_\mu} + {A_\mu}\right) {\psi_\alpha}
+ \frac{1}{2e^2}{{\cal L}_M}(a')\right)
\end{eqnarray}
where ${{\cal L}_M}$ is the Maxwell Lagrangian
and ${a'_\mu}={a_\mu}+{A_\mu}$.
We are left with a Lagrangian with a matter field
which carries both spin
and charge, regardless of whether or not $a'_\mu$ is confining.

In principle, there is another
strongly-coupled phase which is possible -- the Higgs phase --
in which, again, there are no massless gauge
bosons and no free slave particles. In fact
\cite{Osterwalder78,Fradkin79}, the Higgs and confining phases
are not distinct if the Higgs field has gauge charge $1$.
A condensate of slave bosons would be such a phase;
it is a Fermi liquid phase with spin-charge confinement.
If the Higgs field has higher charge -- such as
a composite Higgs formed by a pair of auxiliary
fermions -- then there is a distinct Higgs phase,
but it is still true that there are no massless gauge
bosons and no free slave particles. According to Wen \cite{Wen91},
the short-ranged RVB state \cite{Kivelson87,Rokhsar88}
belongs to such a phase; it is superficially similar
to the phase obtained by condensing $\Phi_\rho$ and
$\Phi_\sigma$ in (\ref{eqn:master}).
However, as reflected in the phase diagram
of Fradkin and Shenker \cite{Fradkin79}, the strong-coupling expansion
is exact in a model with infinite gauge coupling;
it implies that such a phase will not occur. 

Finite-temperature could give us a window into
physics within the confinement length; in
a gauge theory with matter fields, this would
occur via a crossover \cite{Susskind79}.
However, the confinement scale is the lattice scale,
so this would not occur within the physically
relevant temperature regime.

We circumvented the difficulties associated
with a strongly-coupled gauge field by using a dual
representation of the slave particle currents;
in this dual representation, the constraint is linear and,
hence, soluble.  The approach used here might prove
fruitful in the analysis of other models with a
slave-particle formulation.

I would like to thank S. Chakravarty, C. Chamon,
E. Fradkin, Y.B. Kim, and especially C. Mudry and
T. Senthil for discussions;
and S. Kivelson for pointing out reference \cite{LeeDH99}.

{\vskip -0.7 cm}

\end{document}